\documentstyle[preprint,aps]{revtex}
\tighten
\begin{document}
\draft
\preprint{\hfill NCL94--TP4}
\title{ A cosmological no--hair theorem}
\author{Chris M. Chambers and Ian G. Moss}
\address{
Department of Physics, University of Newcastle Upon Tyne, NE1 7RU U.K.
}
\date{March 1994}
\maketitle
\begin{abstract}
A generalisation of Price's theorem is given for application to
Inflationary Cosmologies. Namely, we show that on a Schwarzschild--de
Sitter background there are no static solutions to the wave or
gravitational perturbation equations for modes with angular
momentum greater than their intrinsic spin.
\end{abstract}
\narrowtext

A most intriguing feature of gravitational collapse is the way in which
the final state seems to be characterised by only a few parameters.
This is the phenomenon that John Wheeler described as the loss of
hair by the black hole \cite{ruffini} and forms the content of the
no--hair conjecture.

Implicit in the no--hair conjecture is the idea that gravitational
collapse reaches a stationary state. This is supported by a combination
of numerical and perturbative calculations. The part of the no--hair
conjecture that has been proved rigourously consists of uniqueness
theorems for stationary black hole solutions.

Early work on the non--rotating case can be further sub--divided into
global results \cite{israel} and the results summarised by Price's
theorem \cite{price}:

The only static solutions to the pure massless wave equations with spin
$s=0,\case1/2,1$ or to the gravitational perturbation equations on a
spherically symmetrical black hole background have angular momentum
less than $s$.

The implication is that all of the other modes are radiated away during
the collapse and that the event horizon is characterised by the
constants of integration for the remaining modes. For pure gravity
these are the mass perturbation and the angular momentum of the hole.
(The initial angular momentum being zero).

In this letter we will present a generalisation of Price's theorem for
black holes in asymptotically de Sitter spacetimes. These black holes
have been of interest recently because there are situations in which
they form naked singularities \cite{mellor,kastor,brill,chambers}.
However, the overriding reason for looking at no--hair theorems with a
cosmological constant is because of their role in Inflationary models
of the early universe.

The cosmic no--hair conjecture states that in the presence of a
cosmological constant the universe evolves into a de Sitter universe
\cite{gibbons,hawking}. This would imply that Inflation is a natural
phenomenon that can explain the isotropy and homogeneity seen in the
universe on large scales. In point of fact this simple version of the
cosmic no-hair conjecture is violated simply by generalising the
Schwarzschild spacetime to include a cosmological constant
\cite{kottler}. Valid cosmological no--hair theorems can be obtained
however for the homogeneous universe \cite{wald,jensen} or else in the
vicinity of an infinitely future extendable worldline
\cite{starobinskii}.

It is also possible to derive global uniqueness theorems for stationary
solutions to the Einstein equations for gravity with a cosmological
constant \cite{boucher,boucher1}. What has not been achieved so far is
a theorem of this kind with weak enough conditions at infinity that
would allow black holes and generalise the uniqueness theorems for
static solutions in asymptotically flat spacetimes \cite{israel}.

If the earliest stages of our universe were very chaotic then parts
would have been collapsing under the influence of gravity whilst others
were expanding \cite{linde,maeda,shiromizu,shibata}. It would be
desirable to understand how the formation of black holes would affect
the inflation going on around early episodes of gravitational
collapse.

We will prove the generalisation of Price's theorem first and then
discuss the approach to de Sitter space outside the hole:

The only static solutions to the pure massless wave equations with spin
$s=0,\case1/2,1$ or to the gravitational perturbation equations on a
Schwarzschild--de Sitter background have angular momentum less than
$s$.

The proof requires the explicit forms of the wave equations on the
Schwarzschild--de Sitter background. The background metric is a vacuum
solution to the Einstein equations with a positive cosmological
constant $\Lambda$,
\begin{equation}
ds^2=-r^{-2}\Delta dt^2+r^2\Delta^{-1} dr^2
+r^2(d\theta^2+\sin^2\theta d\phi^2),
\end{equation}
where
\begin{equation}
\Delta=r^2-2Mr-\case1/3\Lambda r^4.
\end{equation}
There are two horizons where $\Delta$ vanishes, a black hole event
horizon at $r=r_2$ and a cosmological horizon at $r=r_1$. The metric
approaches de Sitter space as $r\to\infty$.

Fortuitously, all of the relevant equations on this background now
exist in the literature \cite{mellor,chambers}. We begin with the pure
spin $s$ wave equations. In SL(2,C) spinor notation,
\begin{equation}
\nabla_{AA'}\Psi^{AB\dots C}=0.
\end{equation}
The field has $2s$ symmetrised indices and therefore $2s+1$ independent
components, which we label $\Psi_n$, $n=-s,\dots,s$.

We take,
\begin{equation}
\Psi_n=r^{-s}R_n(j,\omega;r)S_n(j,m;\theta)e^{(im\phi-i\omega t)},
\end{equation}
where the angular functions are spin $n$ eigenfunctions of the
Laplacian on the sphere with eigenvalues $j(j+1)$ and half integer
angular momenta $j$.

The equations for the radial functions on a Kerr--de Sitter background
were analysed recently in ref. \cite{chambers}. In the
Schwarzschild--de Sitter limit they are
\begin{equation}
\left({\cal D}^\dagger_{-n/2}\Delta{\cal D}_{n/2}
+2(2n-1)i\omega r +2(2n-1)(n-1)(Mr^{-1}-\case1/3\Lambda r^2)\right)R_n
=\lambda_n R_n,\label{radiala}
\end{equation}
for $1-s\le n\le s$ and
\begin{equation}
\left({\cal D}_{n/2}\Delta{\cal D}^\dagger_{-n/2}
-2(2n+1)i\omega r +2(2n+1)(n+1)(Mr^{-1}-\case1/3\Lambda r^2)\right)R_n
=\lambda_{-n} R_n,\label{radialb}
\end{equation}
for $-s\le n\le s-1$, with $\lambda_n=(j+n)(j-n+1)$. For $s=0$, which
is excluded by the inequalities,
\begin{equation}
\left({\cal D}_0\Delta{\cal D}^\dagger_0
-2i\omega r \right)R_0=\lambda_0 R_0,
\end{equation}
The operator ${\cal D}_n$ is a radial derivative,
\begin{equation}
{\cal D}_n=\partial_r-{i\omega r^2\over \Delta}+n{\Delta'\over\Delta}.
\end{equation}
(In Kerr--de Sitter the mass term on the left of equations
\ref{radiala} and \ref{radialb} mixes radial and angular modes.
Therefore the examples quoted in ref. \cite{chambers} were all cases
where this term vanished. We have also shifted the subscript on $\Psi$
by $s$.)

In the static case we have $\omega=0$. The equations have regular
singular points at both horizons $r=r_1$ and $r=r_2$ where $\Delta$
vanishes. At these singular points $R_n=O(\Delta^{\pm n/2})$.

Suppose that equation \ref{radiala} has a solution that is regular at
both $r=r_1$ and $r=r_2$. If we multiply the equation by $R_n^\dagger$
and integrate  between the two horizons then,
\begin{equation}
\int_{r_2}^{r_1}\left\{\Delta ({\cal D}_{n/2}R_n)^\dagger
({\cal D}_{n/2}R_n)+V(r) R_n^\dagger R_n\right\}dr
-\left[\Delta R_n^\dagger{\cal D}_{n/2} R_n\right]_{r_2}^{r_1}=0,
\end{equation}
where
\begin{equation}
V(r)=(2n-1)(n-1)(r^{-2}\Delta+\Lambda r^2)+\epsilon(j,n)
\end{equation}
and
\begin{equation}
\epsilon(j,n)=\lambda_n-(2n-1)(n-1)
\end{equation}

The boundary term vanishes for the regular solutions. If $j\ge s>0$
then $\lambda_n\ge s+n$ and $V(r)\ge 0$ for $n=0,\case1/2,1$. If $s=0$
then $V(r)=j(j+1)\ge0$. These are only consistent if $R_n$ vanishes
identically and therefore there are no solutions for $j\ge s$ and
$n\ge0$. The same conclusion follows for negative $n$ from equation
\ref{radialb}. This is the result that was required.

If $s=1$, for example, then only $\Psi_0$ can have a static value. The
solution for the radial modes with $j=0$ is $R_0=Q/r$, where $Q$ is a
constant. This represents the radial electric field of a point charge.

The perturbed Einstein equations are different from the pure spin $2$
wave equation. We follow the discussion in ref. \cite{chandrasekhar},
and decompose the metric perturbations into polar and axial
perturbations depending on their behaviour under $\phi\to-\phi$. For
the polar perturbations
\begin{eqnarray}
\delta g_{rr}&=&2r^2\Delta^{-1} L(r) P_l(\theta),\\
\delta g_{\theta\theta}
&=&2 r^2 (T(r)+U(r) \partial_\theta\partial_\theta) P_l(\theta),\\
\delta g_{\phi\phi}
&=&2 r^2 \sin^2\theta(T(r)+U(r) \cot\theta\partial_\theta)
P_l(\theta).
\end{eqnarray}
All of the functions $T(r)$, $L(r)$ and $U(r)$ are related by the
Einstein equations to a single function,
\begin{equation}
Z_+(r)=rU(r)-r^2(\lambda_2 r+6M)^{-1}(2L(r)+\lambda_2
U(r))\label{defz}.
\end{equation}

For the axial perturbations we set
\begin{eqnarray}
\delta g_{r\phi}&=&
r^2\Delta^{-1}A(r)\sin\theta\partial_\theta P_l(\theta)\\
\delta g_{\theta\phi}&=&
r^2\Delta^{-1}B(r)(\sin\theta\partial^2_\theta
-\cos\theta\partial_\theta)P_l(\theta)
\end{eqnarray}
and define $Z_-(r)=(A(r)-B(r)')/r$. Again, $A(r)$ and $B(r)$ can be
found given $Z_-(r)$.

The perturbed Einstein equations for a charged black hole--de Sitter
background were derived  in ref. \cite{mellor}. With the charge set to
zero these reduce to
\begin{equation}
-{d^2 Z_\pm\over d r^{*2}}+V_\pm(r)Z_\pm=\omega^2 Z_\pm
\end{equation}
where $dr^*=r^2 dr/\Delta$. The potentials are given by
\begin{equation}
V_\pm(r)=\pm 6M \partial_{r^*}f+36 M^2f^2+\lambda_2(\lambda_2+2)f
\end{equation}
where,
\begin{equation}
f={\Delta\over (\lambda_2 r+6M)r^3}.
\end{equation}
The equations are identical in form to the equations that are obtained
for $\Lambda=0$. The cosmological constant appears only in $\Delta$.

It is possible to transform these equations into a similar form to the
previous set. For $\omega=0$ define
\begin{equation}
R_\pm={r^3\over\Delta}
\left(V_\pm(r)Z_\pm + W_\pm(r)\partial_{r^*} Z_\pm\right)\label{zed}
\end{equation}
with
\begin{equation}
W_\pm(r)=-\partial_{r^*}\ln f \mp 6Mf.
\end{equation}
The equations for $R_\pm$ are then
\begin{eqnarray}
\left({\cal D}^\dagger_{-1}\Delta{\cal D}_1
-2\Lambda r^2)\right)R_+
&=&\lambda_2 R_+,\\
\left({\cal D}_{-1}\Delta{\cal D}^\dagger_1
-2\Lambda r^2)\right)R_-
&=&\lambda_2 R_-.
\end{eqnarray}

These are also the equations for the $n=\pm2$ components of the Weyl
tensor that were derived in ref. \cite{chambers}. Proceeding as before,
we find that there are no solutions for $R_\pm$ that are regular on
both horizons apart from  $R_\pm=0$. Setting $R_\pm=0$ (see \ref{zed})
leads to regular solutions for $Z_\pm$ only in the case where $j=1$ and
logarithmically divergent solutions (sufficient for regular $L$ and
$U$) when $j=0$. Therefore the theorem is proven.

The way in which the spacetime around a collapsing star settles down to
the static spacetime is somewhat different from the situation in flat
space. The Penrose diagram in fig. \ref{fig1} shows the collapse of a
spherical star to form a black hole. The metric has been continued
beyond each horizon by introducing Kruskal coordinates. Null
coordinates are defined by $u=t-r^*$ and $v=t+r^*$, where $dr^*=r^2
dr/\Delta$, and the Kruskal coordinates are
\begin{eqnarray}
U_1&=&-e^{-\kappa_1 u},\hbox{~~~~~and~~~}V_1=e^{\kappa_1 v}\\
U_2&=&-e^{-\kappa_2 u},\hbox{~~~~~and~~~}V_2=e^{\kappa_2 v}
\end{eqnarray}
where $\kappa_2$ is the surface gravity of the event horizon.

The values of various field components can be taken to be given on the
stellar surface and then will propagate into the exterior spacetime.
The surface of the star  begins to collapse at $u=u_0$ and lies close
to a null surface $v=v_0$ as it approaches the event horizon. In this
time--dependent problem we let $\Phi(j,m,n;r,t)$ be a field component
with fixed angular eigenvalues $j$ and $m$.

Each mode of the field should remain analytic in the Kruskal
coordinates, and therefore on the surface of the star
\begin{equation}
\Phi\sim \psi_0+\psi_1 e^{-\kappa_2 u}.
\end{equation}
The only modes which have $\psi_0$ non--zero are those with $j<s$. In
flat space the constant modes cannot reach infinity. In de Sitter space
the constant modes propagate all the way to the cosmological horizon.

The surface of the star and the past cosmological horizon ${\cal S_-}$
form the initial data surfaces. If the star is removed the complete
black--hole de Sitter spacetime has a past event horizon ${\cal H_-}$
as well as a future event horizon. Close to the horizons the radial
modes with frequency $\omega$ become plane waves $R_n\sim r^{-1}e^{\pm
i\omega r^*}$. We can define separate reflection and transmission
amplitudes $R(\omega)$ and $T(\omega)$ for modes which propagate from
the past event horizon, denoted by $\rightarrow$ and for modes which
propagate from the past cosmological horizon, denoted by $\leftarrow$.

The initial data corresponds to specifying two functions $F(u)$ and
$G(v)$ on ${\cal H_-}$ and ${\cal S_-}$. These are analytic in $U_2$
and $V_1^{-1}$, the Kruskal coordinates that vanish near ${\cal H_+}$
and ${\cal S_+}$ respectively,
\begin{eqnarray}
F(u)&&\sim F_0+F_1 e^{-\kappa_2 u}\hbox{~~~~as~}u\to\infty\\
G(v)&&\sim G_0+G_1 e^{-\kappa_1 v}\hbox{~~~~as~}v\to\infty.
\end{eqnarray}
The modes that are represented by these expansions have frequencies
$-i\kappa_2$ and $-i\kappa_1$ as well as zero. Therefore after
transmission to the future horizons they become,
\begin{eqnarray}
r\,\Phi(\infty,v)&&\sim (1+\overrightarrow{R(0)})F_0
+\overleftarrow{T(0)}G_0
+\overrightarrow{R(-i\kappa_2)}F_1e^{-\kappa_2v}
+\overleftarrow{T(-i\kappa_1)}G_1e^{-\kappa_1v},\\
r\,\Phi(u,\infty)&&\sim (1+\overleftarrow{R(0)})G_0
+\overrightarrow{T(0)}F_0
+\overleftarrow{R(-i\kappa_1)}G_1e^{-\kappa_1u}
+\overrightarrow{T(-i\kappa_2)}F_1e^{-\kappa_2u}.
\end{eqnarray}
In the asymptotically flat case the corresponding transmission
amplitudes are of order $\omega^{j+1}$ for small omega and the constant
modes are trapped. The present case is much more similar to the
scattering in the interior region of a charged black hole
\cite{chandrasekhar}, from which we can deduce that
\begin{equation}
\overleftarrow{T(0)}\ne 0\hbox{,~~~}
\overrightarrow{T(0)}\ne 0\hbox{,~~~}
\overleftarrow{T(-i\kappa_1)}=\overrightarrow{T(-i\kappa_2)}=0.
\end{equation}
In other words, the constant modes are transmitted to the cosmological
horizon but the exponentially decaying modes are not (compare ref.
\cite{brady}).

Finally, the radiation emitted during the collapse of the star
propagates through the future cosmological horizon and out to infinity.
This radiation lies principally between two retarded times $u_0$ and
$u_1$. In the vicinity of future infinity the $r$ coordinate is
time--like.  It can be related to the Robertson--Walker (k=1)
cosmological time coordinate $\tau$ and radius $\chi$ by
$r=\alpha\cosh(\tau/\alpha)\sin\chi$ where $\alpha^2=3/\Lambda$.
Asymptotically, the modes are bounded by $1/r$ and therefore the
amplitudes seen by an observer at fixed $\chi$ decrease exponentially
in $\tau$ with a decay timescale $\alpha$.

In Inflationary models the cosmological constant is present for only a
limited period but numerical studies of spherically symmetric and
axisymmetric models \cite{shiromizu,shibata} indicate that there is
enough time for massive inhomogeneities to collapse and form stable
black holes before the inflation comes to an end. Thus it seems
possible to have a picture of the early universe in which many
inhomogeneities separated by distances comparable to the cosmological
horizon scale were collapsing into black holes whilst other regions of
the universe were inflating \cite{linde}. Radiation from the collapsing
material would mostly have been damped away by the exponential
expansion in the surrounding spacetime before the end of the
inflationary period. Observers outside of the black holes whose
worldlines extended far into the future would have seen a universe that
increasingly appeared to be de Sitter.

\acknowledgments
We are grateful to Patrick Brady for discussing some of the issues
raised here. C. Chambers is supported by E. P. S. R. C. of Great
Britain.

\begin{figure}
\caption{Penrose diagram for a star collapsing in a de Sitter universe.
The collapse starts at retarded time $u_0$ and the data approaches the
asymptotic form by $u_1$. The diagram extends beyond the edges of the
figure.\label{fig1}}
\end{figure}
\begin{figure}
\caption{Penrose diagram of de Sitter spacetime with many distinct
collapsing lumps and their radiation fields. Most observers see a
universe that approaches the de Sitter universe
asymptotically.\label{fig2}}
\end{figure}

\end{document}